# Ultralow Threshold Polariton Condensate in a Monolayer Semiconductor Microcavity at Room Temperature

Jiaxin Zhao[1,#], Rui Su[1,#], Antonio Fieramosca[1], Weijie Zhao[1], Wei Du[1], Xue Liu[1], Carole Diederichs[2,3], Daniele Sanvitto[4], Timothy C. H. Liew[1,2], Qihua Xiong[1,2,5,*]

[1]Division of Physics and Applied Physics, School of Physical and Mathematical Sciences, Nanyang Technological University, Singapore 637371

[2]MajuLab, International Joint Research Unit UMI 3654, CNRS, Université Côte d'Azur, Sorbonne Université, National University of Singapore, Nanyang Technological University, Singapore

[3]Laboratoire Pierre Aigrain, Département de physique de l'ENS, Ecole Normale Supérieure, PSL Research University, Université Paris Diderot, Sorbonne Paris Cité, Sorbonne Universités, UPMC Univ. Paris 06, CNRS, 75005 Paris, France

[4]CNR NANOTEC Institute of Nanotechnology, via Monteroni, 73100 Lecce, Italy

[5]State Key Laboratory of Low-Dimensional Quantum Physics and Department of Physics, Tsinghua University, Beijing, China

#These authors contributed equally to this work.

*To whom correspondence should be addressed. Email: qihua@ntu.edu.sg

**Abstract**

Atomically thin transition-metal dichalcogenides (TMDs) possess valley-dependent functionalities that are usually available only at crogenic temperatures, constrained by various valley depolarization scatterings. The formation of exciton polaritons (EPs) by coherently superimposing excitons and microcavity photons potentially harnesses the valley-polarized polariton-polariton interactions for novel valleytronics devices. Robust EPs have been demonstrated at room temperature in TMDs microcavity, however, the coherent polariton lasing and condensation remain elusive. Herein, we demonstrate for the first time the realization of EP condensation in a TMD microcavity at room temperature. The continuous-wave pumped EP condensation and lasing with ultralow thresholds (~0.5 nW) is evidenced by the macroscopic occupation of the ground state, that undergoes a nonlinear increase of the emission and a continuous blueshift, a build-up of spatial coherence, and a detuning-controlled threshold. Our work presents a critically important step towards exploiting nonlinear polariton–polariton interactions and polaritonic devices with valley functionality at room temperature.

Material excitations, giving rise to electron-hole bound states called excitons, and photons confined to extremely small volumes in optical microcavities, can give rise to new bosonic quasi-particles called polaritons once in the strong coupling regime[1]. Owing to the advantages of the light-mass of photons and strong nonlinearities of excitons, polaritons can condense, via stimulated scattering[2-4], to a single quantum state even at room temperature[5]. This brings quantum physical phenomena to a macroscopic scale and creates great opportunities for developing novel all-optical devices, for instance, low threshold polariton lasers[6-8], switches and modulators[4,9], polariton lattices[10-13] and networks[14,15] that exploit the great polariton potentiality. Particularly, polariton condensates or, equivalently, polariton lasers, are very interesting devices since they are exempt from the requirement of electronic population inversion, and thus exhibit ultralow threshold lasing. Although polariton condensation has been mainly observed at cryogenic temperatures in CdTe[2] and GaAs[16] microcavity systems, substantial progresses have led to room-temperature operation based on various inorganic[5,17-19] and organic materials[20-25]. The materials that can realize robust polaritonic phenomena at high temperature, by using relatively simple semiconductor fabrication technology, are still crucial for practical polaritonic applications[26].

Recently, two-dimensional group-VI TMDs with direct bandgaps and large oscillator strengths emerged as promising candidates for various applications in novel photonics, optoelectronics and valleytronic devices[27]. Due to the reduced dimensionality and low dielectric screening in TMDs monolayer, their electron-hole interactions are extremely strong with substantial exciton binding energies on the order of several hundred meV[28]. In addition, owing to crystal inversion symmetry breaking and strong spin-orbit coupling in the monolayer TMDs, the excitons at the nonequivalent *K* and *K*' valleys can be coherently and selectively created, which offer new opportunities to explore quantum control of valley pseudospin[27,29-33]. When TMDs are integrated with optical cavities, such as planar microcavities[34], photonic crystals[35] and plasmonic nanocavities[36], robust polaritons have been realized even at room temperature. Furthermore, the spin-valley locking of polaritons paves the way towards breakthroughs in optically manipulating electronic charge and spin degrees of freedom at room temperature. Indeed, in contrast with respect to the bare exciton in TMD, that undergoes a fast intervalley scattering which limits the observation of valley degree of freedom at cryogenic temperature, exciton-polaritons show a persisting valley polarization degree, resulting from the enhanced radiative decay before valley depolarization[30,31]. However, constrained by the limited material gain volume, short microcavity lifetimes, high photonic losses, and broad exciton linewidths, the realization of collective quantum effects, such as condensation or superfluidity, remains elusive. These nonlinear phenomena are major milestones towards the exploitation of functional polaritonic devices with valley degrees of freedom at room temperature in TMDs. In this work, we report for the first time the long-sought polariton condensation based on the hybrid structure of monolayer $WS_2$ in a Fabry-Perot microcavity. We observe the characteristic splitting, indicating the strong coupling regime between neutral excitons and cavity photons, with a Rabi splitting of ~37 meV at room temperature. The continuous wave pumped polariton lasing behavior with an ultralow threshold is unambiguously demonstrated, along with a continuous blueshift beyond threshold, a spatial coherence and an exciton fraction controlled threshold.

To achieve strong coupling between excitons and photons, we incorporate an exfoliated monolayer $WS_2$ into an all-dielectric planar microcavity. The planar Fabry-Perot microcavity with two dielectric distributed-Bragg-reflector (DBR) mirrors used in this work is schematically shown in Figure 1a. The bottom and top DBR mirrors present a near-unity high reflectivity (~99.99%) in the stop band (see Methods and extended data Figure S1). We use the mechanical exfoliation method to obtain high-quality monolayer flakes from bulk crystals, which are then transferred onto the bottom DBR with a silicon dioxide ($SiO_2$) spacer. The optical image of the monolayer area is shown in the supplementary information Figure S2. Steady-state absorption and photoluminescence measurements are used to confirm the thickness and quality of the exfoliated monolayer $WS_2$, shown in Figure 1b. A strong A exciton peak is observed, which is related to the direct optical transition at the *K/K'* point in the Brillouin zone[28]. The photoluminescence spectrum reveals emissions from neutral A excitons and negatively charged excitons (trions) with full widths at half maximum (FWHM) at room temperature of 27 and 53 meV, respectively. We utilize two $SiO_2$ spacer layers to form a sandwich structure around the monolayer $WS_2$ and the thickness of the $SiO_2$ spacer layer is controlled to achieve a microcavity resonance around the neutral A exciton peak at normal incidence. To ensure the strongest light-matter interaction, the monolayer sample is positioned in the middle of the planar microcavity to coincide with the field maximum of the cavity mode. A white-light reflectivity measurement of the empty cavity is presented in supplementary information Figure S3, showing a FWHM of ~0.95 meV and thus indicating a quality factor of ~2100. After embedding the monolayer into the planar cavity, the ground-state emission of WS$_2$ microcavity shifts to ~1.965 eV with a FWHM of ~2.9 meV.

The realization of the strong coupling between excitons and cavity modes in our TMD microcavity is confirmed by angle-resolved reflectivity measurements at room temperature. As shown in Figure 2a, we observe the anti-crossing dispersion of the lower-polariton branch (LPB) and upper-polariton branch (UPB) as a function of the incident angle of light. When the angle increases, the energy of LPB increases towards the exciton energy and is almost invisible for large angles. In contrast, the dispersion of the UPB vanishes at small incident emission angles. These two polariton modes exhibit a clear anti-crossing feature with a Rabi splitting of ~ 37 meV, indicated by the black arrow in Figure 2a. The Rabi splitting is larger than the sum of the FWHMs of the cavity mode and exciton resonance, confirming that the system is well situated in the strong coupling regime. Furthermore, we fit the polariton dispersions by using a coupled oscillator model (more details about the model are shown in the methods section)[4,35], and the calculated polariton dispersions (dashed blue curves) are in good agreement with our experimental data shown in Figure 2a. Here we approximately consider the dispersion of the cavity mode as a parabola-like function, which is supported by reflectivity measurements of the bare cavity (supplementary information Figure S3). The parabolic dispersion of the cavity mode and flat band of excitons (E$_{ex}$ = 2.010 eV) are shown as gray and red dashed curves, respectively. From the fitting of optical measurement data, we extract a negative detuning of $\Delta = E_{ph} - E_{ex} = -34\ meV$, where $E_{ph}$ and $E_{ex}$ are the cavity mode energy and exciton energy, respectively.

To investigate the effect of polariton population, the TMD microcavity is non-resonantly pumped by a 532 nm continuous-wave (CW) laser at room temperature. As schematically illustrated in Figure 1b (inset), the scattering channels between polaritons and the exciton reservoir, and/or the scattering among polaritons further relax the majority of carriers to the ground state of the LPB. Under sufficient polariton density, polariton lasing is triggered by stimulated polariton-polariton scattering, leading to a macroscopic occupation of the polariton ground state as shown in Figure 2b. The inset in Figure 2b displays the angular distribution of the spectrally integrated polariton emission above the threshold where polaritons massively occupy the ground state.

The photoluminescence spectrum at a pump power of ~70 nW (far above threshold) is well fitted with a narrow Lorentzian peak (~1 meV) and a broad background emission, which corresponds to the polariton ground-state emission and non-condensed background emission from the LPB respectively, as shown in Figure 2c. In the inset of Figure 2c, the power-dependent intensity of the narrow ground-state polariton emission extracted from the fitted photoluminescence spectra, shows a typical lasing behavior. A nonlinear "kink" that occurs around the lasing threshold (~0.5 nW, corresponding to ~ 0.06 W/cm$^2$) in the input-output curve is a strong signature of the polariton condensation at room temperature. Compared with photonic lasers based on TMDs[37,38], the threshold of our work is much smaller than that reported in vertical-cavity surface-emitting lasers (5 nW, ~ 0.44 W/cm$^2$) based on monolayer WS$_2$ at room temperature[39], monolayer WSe$_2$ lasing (27 nW, ~ 1 W/cm$^2$) based on photonic crystals at low temperatures[37], and monolayer WS$_2$ (5-8 MW/cm$^2$) lasing in a micro-disk resonator at 10 K[38]. But what really distinguish our work from those previously mentioned is that the observed lasing behavior occurs while the system is in the strong coupling regime, and the stimulated scattering, leading to condensation, is the reason for such extremely low threshold[7] of this phenomenon. Moreover, the observed threshold is several orders of magnitude lower than that one reported in continuous-wave pumped polariton lasers based on III-V GaAs[16] or II-VI CdTe[2] quantum well cavities operated at 5-10 K (see Methods and extended data Table 1 and 2). Compared to bulk material, the stronger electron-hole interaction in the monolayer sample can lead to a large binding energy and increase the optical gain[27,28]. Due to the exceptional exciton oscillator strength in TMD monolayer, a lower threshold and smaller emission linewidth of the laser characteristics are expected[40]. It is noteworthy that a relatively high Q factor of the Fabry-Perot microcavity and the quality of the monolayer are crucial to achieve lasing in the strong coupling regime, an effect that wasn't observed in other monolayer microcavities fabricated by plasma enhanced chemical vapor deposition technique. It is important to note that the threshold behavior will be further presented in momentum-space measurements on microcavities with different detuning in the following text.

To further verify the low threshold lasing behavior, we study the linewidth evolution of the polariton lasing peak as a function of the pump power (Figure 2d). Linewidth narrowing is observed as the optical pump power exceeds the lasing threshold. We extract the FWHM of the polariton ground state mode from Lorentzian fittings. High-resolution PL spectra in the spontaneous emission (0.3 nW) and lasing regime (2 nW) are presented in the inset of Figure 2d for comparison. The FWHM of the emission in the spontaneous emission regime is ~1.3 meV, while it decreases to ~1 meV when the pump intensity is beyond the threshold. The power-dependent linewidth narrowing behavior matches well with the intensity evolution in the input-output curve (Figure 2d), which further confirms the observation of polariton condensation and lasing in our monolayer TMD microcavity system.

To further exclude the possibility of photonic lasing, we conducted temperature-dependent experiment shown in supplementary information Figure S4. Strong coupling between neutral exciton and cavity mode can be observed from room temperature down to 20 K. Compared with the cavity mode itself, which displays nearly no temperature-dependence, the ground state polariton emission exhibits a clear continuous blue shift as the temperature deceases (supplementary information Figure S4d), resulted from the blueshifting exciton level. The lasing behavior was also observed in another monolayer $WS_2$ microcavity device (supplementary information Figure S4 and S5), in which at power below the threshold, the emission spectrum is dominated by a broad background emission, while, increasing the pumping power a sharp peak appears at the ground state energy, *i.e.*, when the system crosses the threshold power. This indicates the high reproducibility of the polariton condensation phenomenon in our devices. Furthermore, part of the unsuppressed background visible above threshold is associated to the spontaneous emission signal that is propagating far away from the excitation spot, as can be seen also in the discussions in the supplementary information Figure S6.

The buildup of phase coherence is another defining signature of polariton lasing. In the low-excitation density regime, polaritons are expected to have a short coherence length limited by their thermal de Broglie wavelength ($\lambda=\sqrt{2\pi^2\hbar^2/mk_BT}$). For WS$_2$ polaritons, the thermal de Broglie wavelength below the threshold is around 0.35 µm. When the excitation is above the threshold, long-range correlations, up to the size of the polariton condensate are expected in the condensed phase[2]. We went further to check the spatial coherence by interferometry measurements. To probe the emergence of spatial coherence, the real space emission image of the polariton lasing was captured by a Michelson interferometer[18]. We replace the mirror in one of the arms by a retroreflector to invert the image in a centro-symmetric way. Thus, the fringe contrast is a measure of the phase coherence between points localized at $\mathbf{r}$ and $-\mathbf{r}$ from the center of polariton lasing emission[2,18,22]. The image of the real space emission of polariton and the inverted one are shown in Figure 2e and 2f, respectively. Correlations measured for the high-density regime are displayed in Figure 2g. The interference fringes are identified over an emission area of ~ 3 µm, which is around 8 times the thermal de Broglie wavelength. This unambiguously suggests the emergence of a spatially extended phase coherence associated to polariton lasing.

Attributing to the half-matter counterpart carried by the exciton, the repulsive interaction between exciton-polariton creates a blueshift of the LPB beyond the lasing threshold, for sufficiently high particle density, as illustrated by tracking the energy position of the ground state as a function of the pumping power, as reported in Figure 3. The measurement is carried out by monitoring the system in real space and a power sweep with multiple spectra is shown in Figure 3a. The energy position is extracted by using a Lorentzian fit of the profiles and show a clear and increasing blueshift as pump power is increased, as reported in Figure 3b. The polariton blueshift trend represents a crucial evidence of polariton lasing in our microcavity samples. Note that the same behavior was found in another sample discussed in supplementary information Figure S7.

Finally, to further corroborate the observation of polariton condensation, we have compared the power dependence of three microcavities possessing different detuning values: -66 meV, -83 meV and -98 meV, respectively. The reflectivity measurements of the bare cavity mode showing a reproducible high Q factor ~2,000, and the corresponding exciton polariton dispersion under strong coupling regime, are shown in supplementary information Figure S8. The three samples are measured under the same excitation conditions (pump power ~ 100 nW), collecting the emission in momentum space (See Methods), and the different polariton dispersions are displayed in Figure 4a, 4b, 4c, respectively. Qualitatively, a similar momentum space intensity distribution is observed, where the ground state is massively occupied accompanied by a weak background associated to spontaneous emission, visible along the LPB. A quantitative analysis is obtained by extracting the intensity profiles (integrated over 10 pixels) from the momentum space maps as a function of the pump power. We considered two different regions: the minimum of the dispersion ($\theta = 0°$, blue vertical line in Figure 4a, 4b, 4c) and the uncondensed polariton ($\theta \neq 0°$, gray vertical line in Figure 4a, 4b, 4c). By using a gaussian fit of the profiles, we evaluated the total intensity for the ground state (background), which is shown in Figure 4d, 4e, 4f, in black (gray) filled squares, respectively. Below the threshold both signals grow linearly, while as the system crosses the threshold the background remains linear and the ground state signal starts to grow with a much larger slope, showing a clear kink in the input-output plot. The visibility of the discontinuity in the input-output plot is clearly evident for high negative detuning, while, as the detuning decreases the system starts to behave as a thresholdless microcavity (See more information at supplementary information Figure S9). Furthermore, as expected, as the excitonic fraction increases the lasing threshold decreases, ranging from ~50 nW (around 4.4 W/cm$^2$) at -98 meV detuning (Figure 4f) until ~3 nW (around 0.27 W/cm$^2$) at -66 meV detuning (Figure 4d and its inset). An overall threshold dependence as a function of the detuning is presented in supplementary information Figure S10, including the data shown in Figure 2 for first device with an even smaller detuning. This observed trend represents a further confirmation that polariton condensation is observed in our planar microcavities. Finally, it is noteworthy that under a CW resonant excitation of the LPB, as shown in the extended data of supplementary information Figure S11, the polariton emission shows prominent valley degree of freedom at room temperature. In addition to the pure demonstration of polariton condensation provided in this paper, our samples could allow the realization of valley addressable polariton condensate.

In summary, we have demonstrated the first room-temperature polariton condensation in a TMD based microcavity. Angle-resolved reflectivity spectra show the clear anti-crossing behavior between UPB and LBP at room temperature with a large Rabi splitting of ~37 meV. The polariton lasing behavior is unambiguously evidenced by an exciton fraction dependent ultralow threshold, linewidth narrowing, power dependent blueshift and build up of spatial coherence length. The ultralow threshold of TMD polariton lasing is much lower than any photonic lasing in TMD materials, as well as many other polariton lasers, at room temperature, using conventional inorganic or organic semiconductors as gain media. The realization of ultralow threshold sources with TMDs at room temperature shows great promise for practical valleytronic applications in the strong coupling regime, such as polariton superfluidity, optical spin switches, and polarization bistabilities with valley degree of freedom. Furthermore, our findings also stimulate future studies towards the realization of room-temperature electrically pumped polariton lasing in van der Waals heterostructure devices.

**Methods**

**Experimental structure:** The monolayer TMD microcavity in our experiment consists of a $\lambda_{ex}/2$ cavity layer sandwiched between two DBRs. The bottom DBR was composed of silicon dioxide (SiO$_2$, n=1.478) spacer with 12.5 alternating pairs of titanium dioxide (TiO$_2$, n=2.498) and SiO$_2$ on the commercial wafer. The high-impurity source materials were used to deposit the DBR by an e-beam evaporator (Cello, Ohmiker-50B). We used WS$_2$ crystals (from HQ Graphene) for mechanical exfoliation and the monolayer samples were transferred on bottom DBR through a dry polymer micro-transfer process. For instance, in a typical process we used a blue tape for repeated exfoliation of a piece of bulk crystal and then the tape was stuck onto a PDMS layer (PF-X4, Gel-Pack). The blue tape was removed and the monolayer WS$_2$ was left behind on a PDMS layer. The WS$_2$/ PDMS film was put onto a bottom cavity then the PDMS was removed, leaving monolayer WS$_2$ on bottom cavity. All measurements were implemented under ambient conditions at room temperature. The monolayer WS$_2$ sample was deposited the SiO$_2$ and the top DBR which was composed of 8.5 alternating pairs of TiO$_2$ and SiO$_2$.

**Optical characterization:** The steady-state absorption spectra of the monolayer WS$_2$ on PDMS were measured by using a micro-spectrophotometer at room temperature. Angle resolved reflectivity and PL spectra were measured in a home-built setup with the Fourier imaging configuration. A high numerical aperture 100× microscope objective (NA=0.85) was used in our angle resolved measurement, which covers an angular range of ±58.3°. The emission from the microcavity was collected through the narrow entrance slit of the spectrometer (Horiba iHR550) and finally onto the 2D charge-coupled device (CCD) array (Horiba, Symphony II). The captures of the CCD array show the angular dependence spectra for both reflectivity and PL. For the angle resolved PL measurements, the pump power was 20 μW, far beyond the ultralow threshold. The steady-state PL of the monolayer WS$_2$ and power dependent real space emission measurement were conducted in a confocal spectrometer (Horiba Evolution 800) by using a continuous-wave laser (532 nm). A high magnification ($100\times$) microscope objective with a numerical aperture of 0.9 was used in real space PL measurements. The power dependence measurements reported in Figure 4 were collected by using a modified inverted microscope (Nikon Ti-u) equipped with a 40X microscope objective (Nikon, NA=0.65) and a tube lens with a focal length of 20 cm. The detected signal, taken from the lateral port of the microscope, is directed into a spectrometer (Horiba iHR550) coupled to a CCD (Horiba, Symphony II), in which, the Fourier plane is projected by using three additional lenses. Along the detection line the real plane of the sample is reconstructed and a square spatial selection (~2.5 x 2.5 µm²) is placed around the excitation spot to cut part of the background signal, so that we observe the exciton polariton dispersion and condensate at extremely small excitation power densities as shown in Figure 4.

**Coupled harmonic oscillator model:** The polariton dispersion can be fitted by using a coupled oscillator model:

$$\begin{pmatrix} E_{exc} - i\gamma_{exc} & g \\ g & E_{cav}(\theta) - i\gamma_{cav} \end{pmatrix} \begin{pmatrix} \alpha \\ \beta \end{pmatrix} = E \begin{pmatrix} \alpha \\ \beta \end{pmatrix}$$

Here the cavity mode can be determined as $E_{cav}(\theta) = E_{ph} / \sqrt{1 - (\sin(\theta)/n_{eff})^2}$. $E_{ph}$ is the photon energy (at zero angle) and $n_{eff}$ is the effective refractive index. $E_{exc}$ is the exciton energy, $\gamma_{exc}$ and $\gamma_{cav}$ are the half-widths of the uncoupled exciton and planar microcavity resonances, respectively. $E_{UPB,LPB}$ are the eigenvalues corresponding to the energies of upper and lower polariton modes. $\alpha$ and $\beta$ are the amplitudes of the eigenvectors, while g is exciton-photon coupling strength. In the strong coupling regime, $E_{UPB,LPB}$ is given by:

$$E_{UPB,LPB} = \frac{1}{2}[E_{exc} + E_{cav} - i(\gamma_{cav} + \gamma_{exc})] \pm \sqrt{g^2 + \frac{1}{4}[E_{exc} - E_{cav} + i(\gamma_{cav} - \gamma_{exc})]^2}$$

where $\hbar\Omega_{Rabi} = 2\sqrt{g^2 - (\gamma_{cav} - \gamma_{exc})^2/4}$ is the Rabi splitting at detuning equal to 0 ($\Delta = E_{exc} - E_{cav} = 0$). A non-vanishing Rabi splitting requires $g > |\gamma_{cav} - \gamma_{exc}|/2$; Furthermore, in the strong coupling regime, the two resonances can be separable, i.e., the Rabi splitting is larger than the sum of the half linewidths of the modes: $\hbar\Omega_{Rabi} > \gamma_{cav} + \gamma_{exc}$.

**Acknowledgments**

Q.X. and T.L. gratefully acknowledge the support from the Singapore Ministry of Education via the AcRF Tier 3 Programme "Geometrical Quantum Materials" (MOE2018-T3-1-002). Q.X. acknowledges the Singapore Ministry of Education Tier 2 project (MOE2017-T2-1-040), and the Singapore National Research Foundation via the NRF-ANR project (NRF2017-NRF-ANR005 2D-Chiral) and the Competitive Research Programme "Integrated On-Chip Planar Coherent Light Sources" (CRP21-2018-0092). T.L. gratefully acknowledges the support from the Singapore Ministry of Education via AcRF Tier 2 project (MOE2018-T2-02-068).

**Author contributions:**

Q.X., J.Z, and R.S. conceived the ideas and designed the experiments. J.Z. prepared the monolayer microcavity samples. J.Z., R.S. and A.F. carried out the optical spectroscopy measurements. J. Z., R.S., A.F., D.S., T.L. and C.D. analyzed the data. J.Z., R.S., A.F., T.L. and Q.X. wrote the manuscript with inputs from all authors.

**Competing interests:** The authors declare that they have no competing interests. Data and materials availability: All data needed to evaluate the conclusions in the paper are present in the paper and/or the Supplementary Materials. Additional data related to this paper may be requested from the authors.

**Reference**

1   Weisbuch, C., Nishioka, M., Ishikawa, A. & Arakawa, Y. Observation of the coupled exciton-photon mode splitting in a semiconductor quantum microcavity. *Physical Review Letters* **69**, 3314 (1992).

2   Kasprzak, J. *et al.* Bose–Einstein condensation of exciton polaritons. *Nature* **443**, 409 (2006).

3   Balili, R., Hartwell, V., Snoke, D., Pfeiffer, L. & West, K. Bose-Einstein condensation of microcavity polaritons in a trap. *Science* **316**, 1007-1010 (2007).

4   Deng, H., Haug, H. & Yamamoto, Y. Exciton-polariton bose-einstein condensation. *Reviews of Modern Physics* **82**, 1489 (2010).

5   Christopoulos, S. *et al.* Room-Temperature Polariton Lasing in Semiconductor Microcavities. *Physical Review Letters* **98**, 126405 (2007).

6   Imamog lu, A., Ram, R. J., Pau, S. & Yamamoto, Y. Nonequilibrium condensates and lasers without inversion: Exciton-polariton lasers. *Physical Review A* **53**, 4250-4253 (1996).

7   Deng, H., Weihs, G., Snoke, D., Bloch, J. & Yamamoto, Y. Polariton lasing vs. photon lasing in a semiconductor microcavity. *Proceedings of the National Academy of Sciences* **100**, 15318-15323 (2003).

8   Fraser, M. D., Höfling, S. & Yamamoto, Y. Physics and applications of exciton–polariton lasers. *Nature materials* **15**, 1049 (2016).

9   Amo, A. *et al.* Exciton–polariton spin switches. *Nature Photonics* **4**, 361 (2010).

10  Lai, C. *et al.* Coherent zero-state and π-state in an exciton–polariton condensate array. *Nature* **450**, 529-532 (2007).

11  Jacqmin, T. *et al.* Direct observation of Dirac cones and a flatband in a honeycomb lattice for polaritons. *Physical review letters* **112**, 116402 (2014).

12  Baboux, F. *et al.* Bosonic condensation and disorder-induced localization in a flat band. *Physical review letters* **116**, 066402 (2016).

13  Su, R. *et al.* Observation of exciton polariton condensation in a perovskite lattice at room temperature. *Nature Physics* **16**, 301-306 (2020).

14  Amo, A. *et al.* Collective fluid dynamics of a polariton condensate in a semiconductor microcavity. *Nature* **457**, 291-295 (2009).

15  Berloff, N. G. *et al.* Realizing the classical XY Hamiltonian in polariton simulators. *Nature materials* **16**, 1120-1126 (2017).

16  Bajoni, D. *et al.* Polariton Laser Using Single Micropillar GaAs-GaAlAs Semiconductor Cavities. *Physical Review Letters* **100**, 047401 (2008).

17  Xie, W. *et al.* Room-Temperature Polariton Parametric Scattering Driven by a One-Dimensional Polariton Condensate. *Physical Review Letters* **108**, 166401 (2012).

18  Su, R. *et al.* Room-temperature polariton lasing in all-inorganic perovskite nanoplatelets. *Nano Letters* **17**, 3982-3988 (2017).

19  Su, R. *et al.* Room temperature long-range coherent exciton polariton condensate flow in lead halide perovskites. *Science advances* **4**, eaau0244 (2018).

20  Lanty, G., Brehier, A., Parashkov, R., Lauret, J.-S. & Deleporte, E. Strong exciton–photon coupling at room temperature in microcavities containing two-dimensional layered perovskite compounds. *New Journal of Physics* **10**, 065007 (2008).

21  Kena Cohen, S. & Forrest, S. Room-temperature polariton lasing in an organic single-crystal microcavity. *Nature Photonics* **4**, 371 (2010).

22  Daskalakis, K., Maier, S., Murray, R. & Kena Cohen, S. Nonlinear interactions in an organic polariton condensate. *Nature Materials* **13**, 271 (2014).

23  Lerario, G. *et al.* Room-temperature superfluidity in a polariton condensate. *Nature Physics* **13**, 837 (2017).

24  Wang, J. *et al.* Room temperature coherently coupled exciton–polaritons in two-dimensional organic–inorganic perovskite. *ACS nano* **12**, 8382-8389 (2018).

25  Fieramosca, A. *et al.* Two-dimensional hybrid perovskites sustaining strong polariton interactions at room temperature. *Science advances* **5**, eaav9967 (2019).

26  Sanvitto, D. & Kéna-Cohen, S. The road towards polaritonic devices. *Nature materials* **15**, 1061-1073 (2016).

27  Xu, X., Yao, W., Xiao, D. & Heinz, T. F. Spin and pseudospins in layered transition metal dichalcogenides. *Nature Physics* **10**, 343 (2014).

28  Chernikov, A. *et al.* Exciton binding energy and nonhydrogenic Rydberg series in monolayer $WS_2$. *Physical Review Letters* **113**, 076802 (2014).

29  Zeng, H., Dai, J., Yao, W., Xiao, D. & Cui, X. Valley polarization in $MoS_2$ monolayers by optical pumping. *Nature Nanotechnology* **7**, 490 (2012).

30  Chen, Y.-J., Cain, J. D., Stanev, T. K., Dravid, V. P. & Stern, N. P. Valley-polarized exciton–polaritons in a monolayer semiconductor. *Nature Photonics* **11**, 431 (2017).

31  Sun, Z. *et al.* Optical control of room-temperature valley polaritons. *Nature Photonics* **11**, 491 (2017).

32  Qiu, L., Chakraborty, C., Dhara, S. & Vamivakas, A. Room-temperature valley coherence in a polaritonic system. *Nature communications* **10**, 1513 (2019).

33  Lundt, N. *et al.* Optical valley Hall effect for highly valley-coherent exciton-polaritons in an atomically thin semiconductor. *Nature Nanotechnology* **14**, 770-775 (2019).

34  Liu, X. *et al.* Strong light–matter coupling in two-dimensional atomic crystals. *Nature Photonics* **9**, 30 (2015).

35  Zhang, L., Gogna, R., Burg, W., Tutuc, E. & Deng, H. Photonic-crystal exciton-polaritons in monolayer semiconductors. *Nature Communications* **9**, 713 (2018).

36  Cuadra, J. *et al.* Observation of tunable charged exciton polaritons in hybrid monolayer $WS_2$ plasmonic nanoantenna system. *Nano Letters* **18**, 1777-

1785 (2018).
37   Wu, S. *et al.* Monolayer semiconductor nanocavity lasers with ultralow thresholds. *Nature* **520**, 69 (2015).
38   Ye, Y. *et al.* Monolayer excitonic laser. *Nature Photonics* **9**, 733 (2015).
39   Shang, J. *et al.* Room-temperature 2D semiconductor activated vertical-cavity surface-emitting lasers. *Nature communications* **8**, 543 (2017).
40   Arakawa, Y. & Yariv, A. Quantum well lasers--Gain, spectra, dynamics. *IEEE journal of quantum electronics* **22**, 1887-1899 (1986).

**Figures and Captions**

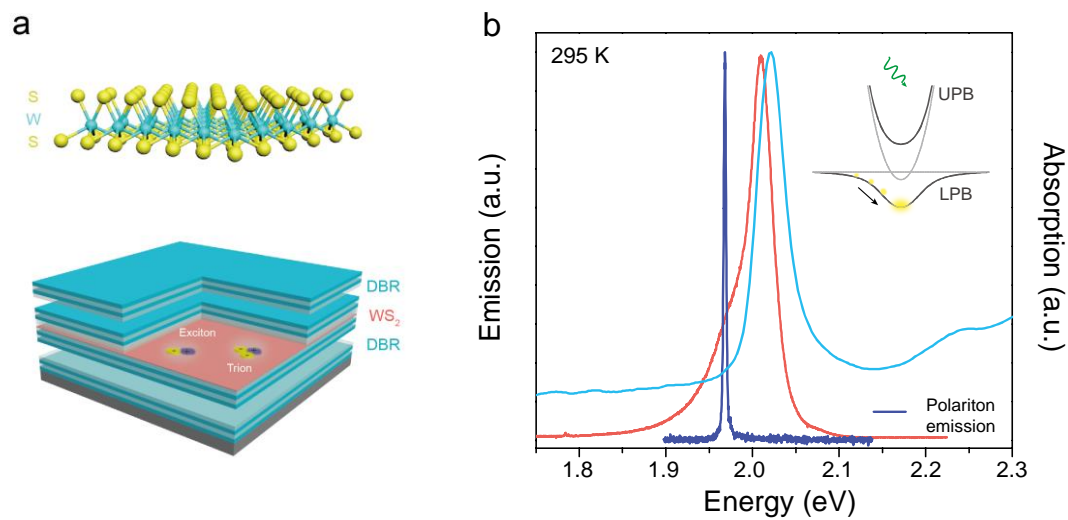

**Figure 1 | Sample structure and monolayer characterization. a,** Schematic diagram of the microcavity structure. This structure consists of 12.5 periods of a bottom distributed DBR ($SiO_2$/$TiO_2$) and 8.5 periods of a top DBR. The cavity layer structure includes a quantum well structure with the monolayer $WS_2$ flake as active material and the $SiO_2$ layers are used as spacer. **b,** monolayer characterization at room temperature. The absorption spectrum (blue curve) of the monolayer $WS_2$ on PDMS (polydimethylsiloxane), showing a strong excitonic peak around 2.02 eV for the A exciton. The photoluminescence spectrum (red curve) of the monolayer $WS_2$ on bottom DBR, showing two emission peaks at ~2.01 eV for neutral A exciton with FWHM of ~27 meV and 1.974 eV (for trion). The ground-state polariton emission (purple curve) of the monolayer $WS_2$ microcavity, showing an emission peak at ~1.968 eV with a FWHM of ~2.9 meV at room temperature. The inset shows the schematics of principles for polariton lasing.

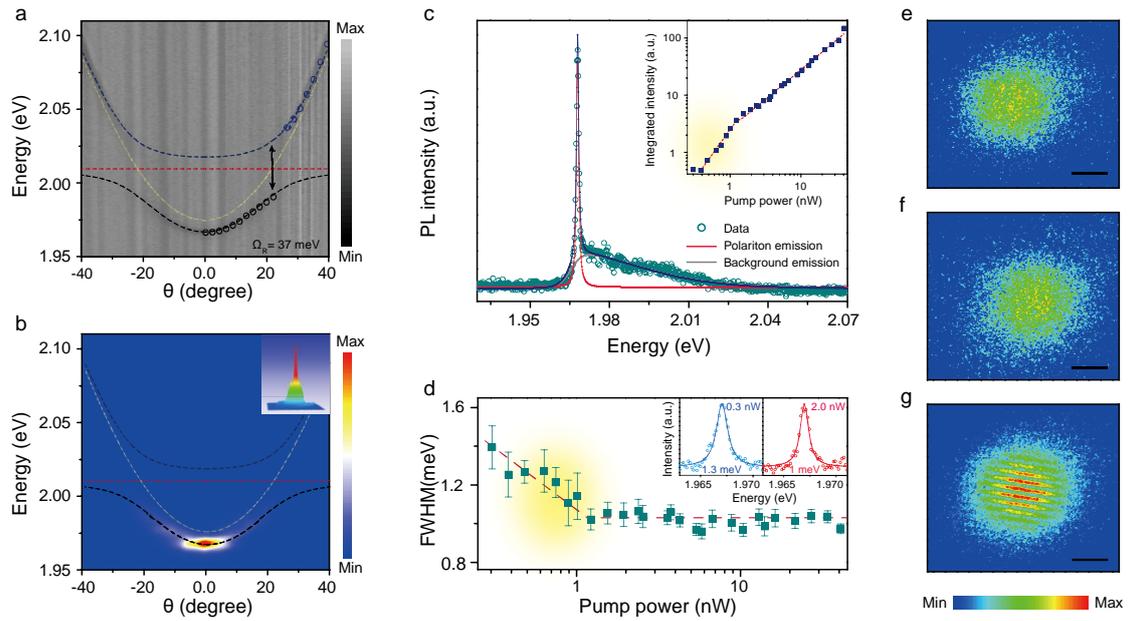

**Figure 2 | Observation of WS$_2$ microcavity polariton lasing at RT. a,** Angle-resolved reflectivity map measured by using a white light source. The dashed blue and black curves show the theoretically fitted dispersions of the UPB and LPB with a Rabi splitting of ~ 37 meV. The dashed red and yellow lines show the uncoupled exciton and the bare cavity mode, which are obtained from a coupled harmonic oscillator model. **b,** Angle-resolved photoluminescence map above threshold. A sharp and intense peak is visible in the ground state, corresponding to the lowest momentum state (inset). **c,** Real space photoluminescence spectrum with pump power of ~ 70 nW. The circles show the measured data, the gray curve is a fitting result to the non-condensate polariton background emission, and the red curve is a fit to the ground-state polariton emission of WS$_2$ cavity emission. The ground-state polariton emission as a function of the pump power (input-output curve) is shown in the inset. **d,** Linewidth as a function of the pump power. Dashed red line is a guide to the eye. Linewidth narrowing of the lasing mode is observed as the system crosses the lasing threshold (the yellow colored area). High-resolution photoluminescence spectra at very low pump power (below the threshold) and high pump power (above the threshold) are shown in the inset. The FWHM of the polariton emission is reduced from ~ 1.3 meV to ~ 1 meV. **e, f** Real space photoluminescence image measured above threshold from the two arms of a Michelson interferometer where a retroreflector is used to flip the image in a centrosymmetric way. **g,**

Interference pattern measured above threshold from the two arms of the interferometer. The scale bar corresponds to 1 μm.

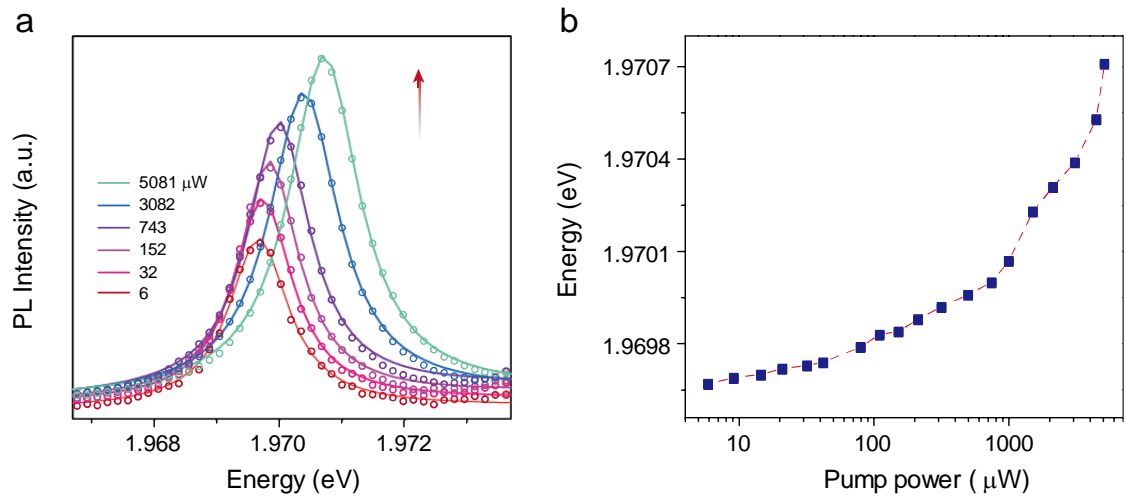

**Figure 3 | Observation of polariton interactions in WS$_2$ microcavity. a,** Real space photoluminescence spectra of the polariton ground state emission as a function of pumping power illustrated in the figure. The colored points indicate the experimental data which are fitted with a Gaussian function, colored solid lines. The gray dashed line is a guide to the eye. **b**, Plot of the energy position of the ground state of the LPB, extracted from the Lorentzian fit reported in (a) for different excitation powers. A clear blueshift is visible.

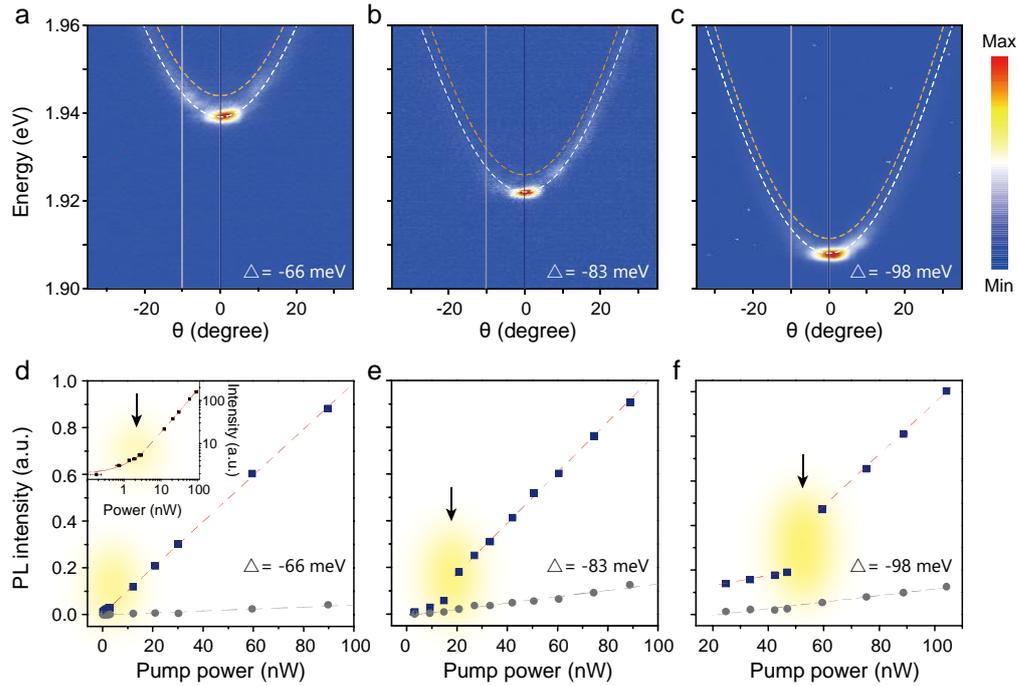

**Figure 4 | Observation of polariton condensation in WS$_2$ microcavity with different detuning. a, b, c,** Above threshold angle-resolved photoluminescence maps of monolayer WS$_2$ microcavity with different detuning ~66 meV (a), ~83 meV (b) and ~98 meV (c). The white (orange) dashed line indicates the LPB (bare cavity mode), respectively. The blue and gray vertical lines indicate the region (averaged over 10 pixels) used to calculate the photoluminescence intensity reported in (d),(e),(f). **d, e, f,** Photoluminescence intensity of polariton ground state ($\theta = 0°$) and background ($\theta \neq 0°$) as a function of pump fluence for different detuning, -66 meV (d), -83 meV (e) and -98 meV (f), respectively. The blue (gray) filled squares correspond to the ground state (background) emission. The threshold region is indicated by the black arrow and yellow colored region while the red dashed line is a linear fit of the data and a guide for the eye to underline the different slopes above threshold.